\documentstyle[12pt]{article}

\newcommand{\EQ}{\begin{equation}}
\newcommand{\EN}{\end{equation}}
\newcommand{\bea}{\begin{eqnarray}}
\newcommand{\ena}{\end{eqnarray}}
\newcommand{\bdis}{\begin{displaymath}}
\newcommand{\edis}{\end{displaymath}}
\newcommand{\vs}[1]{\vspace{#1 mm}}

\renewcommand{\a}{\alpha}
\renewcommand{\b}{\beta}
\renewcommand{\c}{\gamma}
\renewcommand{\d}{\delta}
\renewcommand{\v}{\Delta}
\renewcommand{\o}{\omega}
\renewcommand{\t}{\tau}

\newcommand{\pa}{\partial}

\newcommand{\nn}{\nonumber \\}

\newcommand{\HFP}{H_{FP}}
\newcommand{\tHFP}{{\tilde H}_{FP}}

\newcommand{\hA}{\hat A}
\newcommand{\hp}{\hat \pi}

\newcommand{\limt}{\lim_{\t \rightarrow \infty}}
\newcommand{\itd}{\int \! d^3x}

\newcommand{\igr}{\int \! {\cal D}A}
\newcommand{\dpi}{{\d \over \d A^{in}}}
\newcommand{\hi}{{\hat i}}
\newcommand{\hj}{{\hat j}}

\begin{document}

\topmargin 0pt
\oddsidemargin 5mm

\newcommand{\NP}[1]{Nucl.\ Phys.\ {\bf #1}}
\newcommand{\PL}[1]{Phys.\ Lett.\ {\bf #1}}
\newcommand{\CMP}[1]{Comm.\ Math.\ Phys.\ {\bf #1}}
\newcommand{\PR}[1]{Phys.\ Rev.\ {\bf #1}}
\newcommand{\PRL}[1]{Phys.\ Rev.\ Lett.\ {\bf #1}}
\newcommand{\PREP}[1]{Phys.\ Rep.\ {\bf #1}}
\newcommand{\PTP}[1]{Prog.\ Theor.\ Phys.\ {\bf #1}}
\newcommand{\PTPS}[1]{Prog.\ Theor.\ Phys.\ Suppl.\ {\bf #1}}
\newcommand{\NC}[1]{Nuovo.\ Cim.\ {\bf #1}}
\newcommand{\JPSJ}[1]{J.\ Phys.\ Soc.\ Japan\ {\bf #1}}
\newcommand{\MPL}[1]{Mod.\ Phys.\ Lett.\ {\bf #1}}
\newcommand{\IJMP}[1]{Int.\ Jour.\ Mod.\ Phys.\ {\bf #1}}
\newcommand{\AP}[1]{Ann.\ Phys.\ {\bf #1}}
\newcommand{\RMP}[1]{Rev.\ Mod.\ Phys.\ {\bf #1}}
\newcommand{\PMI}[1]{Publ.\ Math.\ IHES\ {\bf #1}}
\newcommand{\JETP}[1]{Sov.\ Phys.\ J.E.T.P.\ {\bf #1}}
\newcommand{\TOP}[1]{Topology\ {\bf #1}}
\newcommand{\AM}[1]{Ann.\ Math.\ {\bf #1}}
\newcommand{\LMP}[1]{Lett.\ Math.\ Phys.\ {\bf #1}}
\newcommand{\CRASP}[1]{C.R.\ Acad.\ Sci.\ Paris\ {\bf #1}}
\newcommand{\JDG}[1]{J.\ Diff.\ Geom.\ {\bf #1}}
\newcommand{\JSP}[1]{J.\ Stat.\ Phys.\ {\bf #1}}

\begin{titlepage}
\setcounter{page}{0}
\begin{flushright}
NBI-HE-95-26\\
August 1995\\
\end{flushright}

\vs{8}
\begin{center}
{\Large Stochastic Quantization of Matrix Models \\
and Field Theory of Non-Orientable Strings
\footnote{
To appear in Proceedings of the 6th Moscow Quantum Gravity Seminar,
Moscow, June 12-17, 1995.}
}

\vs{15}
{\large Naohito Nakazawa\footnote{On leave of absence from Department
of Physics, Faculty of Science, Shimane University, Matsue 690,
Japan.\quad
e-mail:
nakazawa@nbivax.nbi.dk, nakazawa@ps1.yukawa.kyoto-u.ac.jp}}\\
{\em The Niels Bohr Institute, Blegdamsvej 17, DK-2100 Copenhagen \O,
Denmark \\ }
\end{center}

\vs{8}
\centerline{{\bf{Abstract}}}

In quantizing gravity based on stochastic quantization method,
the stochastic time plays a role of the proper time.
We study 2D and 4D Euclidean quantum gravity in this context.
By applying stochastic quantization method to
real symmetric matrix models, it is shown that
the stochastic process defined by the
Langevin equation in loop space describes the time evolution of the
non-orientable loops which defines non-orientable 2D surfaces.
The corresponding Fokker-Planck
hamiltonian deduces a non-orientable string field theory
at the continuum limit. The strategy, which we have learned in the
example of 2D quantum gravity, is applied to 4D case.
Especially, the
Langevin equation for the stochastic process of
3-geometries is proposed to describe the ( Euclidean ) time
evolution in 4D quantum gravity with Ashtekar's canonical variables.
We present it in both lattice regularized version and the naive
continuum limit.

\end{titlepage}
\newpage
\renewcommand{\thefootnote}{\arabic{footnote}}
\setcounter{footnote}{0}
\centerline{\bf{Introduction}}

String field theory~\cite{SFT} is believed to be the most promising approach
to investigate non-perturbative effect in string theories.
Recently, non-critical string field theories have been proposed
for $c = 0$~\cite{IK}\cite{JR}\cite{Wata}\cite{Na1} and
for $0< c < 1$~\cite{IIKMNS}~\cite{Kos}.
Among these, some~\cite{IK}\cite{Wata}\cite{IIKMNS}\cite{Kos}
are based on the transfer-matrix formalism~\cite{KKMW} in dynamical
triangulation of random surfaces~\cite{DT}.
While some~\cite{JR}\cite{Na1} are derived by using
stochastic quantization method~\cite{PW}.

In the approach by stochastic quantization of matrix models,
one can interpret the stochastic
( fictitious ) time as a Euclidean time coordinate in 2D quantum gravity.
There is also another example of this particular observation,
the proper time interpretation of the
stochastic time, which was found in the course to
study $QCD_4$ in terms of the
Nicoli-Langevin maps~\cite{Ni},
that stochastic quantization of
3D Chern-Simons theory recovers the time
evolution in 4D Euclidean Yang-Mills theory~\cite{CH}.
These facts motivate us to interpret the stochastic time as the
time coordinate in Euclidean 4D quantum gravity~\cite{Na3}.

In this paper,  we illustrate how to apply
stochastic quantization method to real symmetric matrix models~\cite{RS}
and show that it leads to a field theory of non-orientable (non-critical)
strings~\cite{Na1}.
The stochastic process defined by the
Langevin equation in loop space describes the time evolution of the
non-orientable loops on non-orientable 2D surfaces.
The corresponding Fokker-Planck
hamiltonian is a loop space hamiltonian of
non-orientable string field theories. At the equilibrium limit, it deduces
the Virasoro constraint equation for the probability distribution
functional.
The continuum limit of the field theory of discretized non-orientable loops
is taken for the simplest one-matrix case ( $c = 0$ )
and deduces the continuum field theory of
non-orientable strings.

Then we apply the strategy we have learned in 2D case to 4D Euclidean
quantum gravity~\cite{Na3}.
The Langevin equation for 3-geometries is proposed in the Ashtekar's
canonical variables to describe the time evolution in
4D Euclidean quantum gravity in
the sense, that the corresponding Fokker-Planck hamiltonian recovers
the hamiltonian of 4D quantum gravity exactly.
The stochastic time corresponds to
Euclidean time in the temporal gauge,
$N=1$ and $N^i=0$. In this context, 4D quantum gravity is understood
as a stochastic process where the scale of the fluctuation
of \lq \lq triad \rq\rq
is characterized by the curvature at one unit time step before.
The lattice regularization of the approach in 4D Euclidean spacetime
is also presented to play
the same game as what we have done in 2D case with matrix models.

\vs{8}
\centerline{\bf{Stochastic Quantization of Real Symmetric Matrix
Models in Loop Space}}

Let us start with the Langevin equation for one matrix model,
\bea
{\v}M_{ij}(\t)
&=&  - {\pa \over\pa M}S(M)_{ij}(\t) \v\t + \v\xi_{ij}(\t)        \ , \nn
S(M)
&=& - \sum_{\a = 0} {g_\a \over \a + 2} N^{- \a /2}{\rm tr}M^{\a + 2}  \ , \nn
\ena
$M_{ij}$ denotes a real symmetric matrix.
The stochastic time $\t$ is
discretized with the unit time step $\v\t$. We consider the
discretized version of time evolution
$
M_{ij} ( \t+\v\t ) \equiv M_{ij} ( \t ) + \v M_{ij}( \t ) \ ,
$
with the Langevin equation for convenience of
stochastic calculus and for understanding the corresponding
stochastic process precisely.
The discretized stochastic time development with $\v\t$ precisely
corresponds to the
one step deformation in dynamical triangulation in random surfaces.
In the following argument,
the specific form of the action of the matrix model is not relevant.
The correlation of the white noise $\v\xi_{ij}$ is
defined by
\EQ
<\v\xi_{ij}(\t) \v\xi_{kl}(\t)>_\xi
= \v\t \big( \d_{il} \d_{jk} + \d_{ik} \d_{jl} \big)   \ .
\EN
It is uniquely determined\footnote{
For an hermitian matrix $M_{ij}$ in (1),
the nose correlation is
$
<\v\xi_{ij}(\t) \v\xi_{kl}(\t)>_\xi
= 2 \v\t \d_{il} \d_{jk}                        \ .
$
%
} from the requirement that
(1) is transformed covariantly preserving the white noise
correlation (2) invariant under the transformation
$
M \rightarrow U M U^{-1}   \ ,
$
where $U$ denotes orthogonal matrices for the
real symmetric matrix models.

The basic field variables are loop variables
$
\phi_n = {\rm tr}(M^n) N^{-1 - {n\over 2}}   \ .
$
Following to Ito's stochastic calculus~\cite{I}, we evaluate
\bea
\v\phi_n
&\equiv & \phi_n (\t + \v\t) - \phi_n (\t)                  \ , \nn
&=& n {\rm tr}(\v M M^{n-1}) N^{-1 - {n\over 2}}
+ {1\over 2} n \sum_{k=0}^{n-2} {\rm tr}(\v M M^{k} \v M M^{n-k-2})
N^{-1 - {n\over 2}}
+ O(\v\t^{3/2})                          \  .   \nn
\ena
The terms in R.H.S. should be of the order $\v\t$, thus we obtain
\bea
\v\phi_n
&=& \v\t { n\over 2} \big\{ \sum_{k=0}^{n-2}
\phi_k \phi_{n-k-2}
+ (n-1) {1\over N} \phi_{n-2}                 \big\}
+ \v\t\ n \sum_{\a=0} g_\a \phi_{n+\a}   +   \v \zeta_{n-1}    \ ,   \nn
\v\zeta_{n-1}
&\equiv& n {\rm tr}(\v\xi M^{n-1}) N^{-1 - {n\over 2}}     \ .  \nn
\ena
The correlation of the new noise variables appeared in (4) is given by
\EQ
<\v\zeta_{m-1} (\t) \v\zeta_{n-1} (\t)>_\xi
= \v\t {2\over N^2} n m < \phi_{m+n-2} (\t) >_\xi   ,
\EN
The new noise is not a simple white noise but includes the
value of the loop variable itself.
In a practical sense, it might be
tedious to generate the noise variable.
We notice that
$\phi_{m+n-2}(\t)$ in R.H.S. of eq.(5) does not include the white noise
at $\t$ but the series of noises up to
the one step (stochastic time unit $\v\t$) before.
This means that the expectation value in R.H.S. should be defined with
respect to the white noise correlation up to the stochastic time $\t - \v\t$.

We also notice
$
<\v\zeta_n (\t)>_\xi = 0
$
by means of Ito's stochastic calculus.
In the context of SQM approach,
the property of the noise yields
the Schwinger-Dyson equation by assuming the existence of the
equilibrium limit at the infinite stochastic time, or equivalently,
$
\lim_{\t \rightarrow \infty} < \v \phi_n (\t) >_\xi = 0    \  .
$
We have,
\EQ
< {n\over 2} \sum_{k=0}^{n-2}
\phi_k \phi_{n-k-2}
+ {1\over 2}(n-1) {1\over N} \phi_{n-2}
+ n \sum_{\a=0} g_\a \phi_{n+\a}     >_\xi = 0                \ .
\EN

The order of the noise correlation (5), ${1/N^2}$,
realizes the factorization condition in the large N limit.
Therefore we obtain the S-D equation at large N limit for discretized
non-orientable strings.
\EQ
{1\over 2}\sum_{k=0}^{n-2} < \phi_k >_\xi< \phi_{n-k-2} >_\xi
+ \sum_{\a = 0} g_\a < \phi_{n + \a}  >_\xi = 0                \ .
\EN
This shows that the S-D equation for non-orientable strings takes the
same form as that for orientable strings at large N limit.
The correspondence at the large N limit is exact if we define the
corresponding hermitian matrix model by replacing all the couplings
, $g_\a \rightarrow 2g_\a$ in (1).
As a consequence,
the disc amplitude in non-orientable strings is exactly the same as
that in orientable
strings.

The geometrical meaning of the stochastic process described by
the Langevin equation (4) is the following.
The one step stochastic time evolution
of a discretized loop,
$
\phi_n (\t) \rightarrow \phi_n(\t) + \v \phi_n (\t)    \ ,
$
generates the splitting of the loop into two smaller pieces,
$\phi_k$ and
$\phi_{n-k-2}$. The process is described by the first term in R.H.S.
of (4). In a field theoretical sense, it is interpreted as
the annihilation of the loop $\phi_n$ and the
simultaneous pair creation of loops, $\phi_k$ and $\phi_{n-k-2}$.
The first term in R.H.S. of (4) preserves the
orientation of these loops, while the second term, which
is the characteristic term of the
order of ${1\over N}$
for non-orientable interaction,
does not preserve the orientation.
Since the new noise variables in (5),
$\v\zeta_{n-1}$'s, are translated to \lq\lq annihilation" operators in
the corresponding Fokker-Planck hamiltonian,
the factor 2 in the correlation (5) for the
new noise variables comes from the sum of the orientation preserving and
non-preserving merging interactions. Namely, (5) describes the
simultaneous annihilation of two loops
$\phi_{m}$ and $\phi_{n}$ and the creation of a loop $\phi_{m+n-2}$.
The geometrical picture allows us to identify the power \lq\lq $n$" of matrices
in $\phi_n$ to the length of the discretized non-orientable loop $\phi_n$.
We notice that, in each time step, the interaction process decreases the
discretized loop length by the unit \lq\lq 2".
The process which comes from the
original action of matrix models extends the length
of discretized loops.

The definition of the F-P hamiltonian operator
gives the precise definition
of a field theory of second quantized non-orientable strings.
In terms of the expectation value of an observable $O(\phi)$, a
function of $\phi_n$'s, the F-P hamiltonian operator ${\hat H}_{FP}$ is
defined by,
\EQ
<\phi (0)| {\rm e}^{- \t {\hat H}_{FP} } O({\hat \phi})|0>
\equiv <O( \phi_\xi(\t) )>_\xi                  \  .
\EN
In R.H.S., $\phi_\xi(\t)$ denotes the solution of the Langevin equation (4)
with the initial configuration $\phi (0) \neq 0$.
The time evolution of R.H.S. is given by,
\bea
<\v O(\phi(\t))>_\xi
&=& <\sum_m \pa_m O (\phi(\t)) \v \phi_m
+ {1\over 2}\sum_{m,n} \pa_m \pa_n O(\phi(\t)) \v \phi_m \v \phi_n >_\xi
+ O(\v \t^{3/2})         \ ,      \nn
&\equiv& - \v \t < H_{FP}(\t) O(\phi(\t)) >_\xi      \ , \nn
\ena
where $\pa_n \equiv {\pa \over \pa \phi_n}$. By substituting the Langevin
equation (4) and the noise correlation (5) into (9), we obtain
\bea
H_{FP}(\t)
&=& - \sum_{n>0}X_{n} n\pi_n     \ , \nn
X_n
&\equiv&
{1\over N^2} \sum_m m\phi_{m+n-2}\pi_m
+ {1\over 2}\sum_{k=0}^{n-2}\phi_k\phi_{n-k-2}
+ {1\over 2}(n-1) {1\over N} \phi_{n-2}
+ \sum_{\a =0}g_\a \phi_{n + \a}                     \nn  ,
\ena
where
$
\pi_n \equiv {\pa \over\pa \phi_n}     \ .
$
To define the operator formalism corresponding to eq.(8),
we introduce
${\hat \phi}_m$ and ${\hat \pi}_m$ as the creation and the annihilation
operators for the loop with the length $n$, respectively. Then we assume
the commutation relation
$
[ {\hat \pi}_m  ,  {\hat \phi}_n ]
= \d_{mn}                 \ ,
$
and the existence of the vacuum, $|0>$, with
$
{\hat \pi}_m |0> =  <0|{\hat \phi}_m
= 0 \quad {\rm for} \  m > 0      \  .
$
In the representation,
$
<Q| \equiv <0|{\rm e}^{\sum_m Q_m {\hat \pi}_m}  \
$
and
$
|Q> \equiv \Pi_m \d ({\hat \phi}_m - Q_m ) |0>  \ ,
$
the F-P hamiltonian operator
${\hat H}_{FP}$ in (8) is given by replacing
$\phi_m \rightarrow {\hat \phi}_m$, and
$\pi_m \rightarrow {\hat \pi}_m$ in $H_{FP}$ in (10) with the same operator
ordering.

{}From the equality (8), the probability distribution function $P(\phi, \t)$
, which is defined by
$
<O( \phi(\t) )>_\xi
\equiv \int \!\Pi_{n} d\phi_n O(\phi) P(\phi, \t)    \ ,
$
is given by,
\EQ
P(\phi,\t)
= <\phi (0)| {\rm e}^{- \t {\hat H}_{FP} } | \phi >          \  .
\EN
The initial distribution,
$
P(\phi, 0) = \Pi_m \d (\phi_m - \phi_m (0) )     \ ,
$
represents the initial value of the solution of the Langevin equation
(4). Eq.(11) follows the Fokker-Planck equation for the probability
distribution,
\EQ
\v P(\phi,\t)
= + \v\t {\tilde H}_{FP}P(\phi,\t)      \ ,
\EN
where ${\tilde H}_{FP}$ is the adjoint of $H_{FP}$ in (10),
\bea
{\tilde H}_{FP}
&=& - \sum_{n>0} n\pi_n {\tilde X}_{n}      \ , \nn
{\tilde X}_n
\equiv
&-& {1\over N^2} \sum_m m\pi_m \phi_{m+n-2}
+ {1\over 2}\sum_{k=0}^{n-2}\phi_k\phi_{n-k-2}
+ {1\over 2}(n-1) {1\over N} \phi_{n-2}                 \nn
&+& \sum_{\a =0}g_\a \phi_{n + \a}
                  \  .        \nn
\ena
The remarkable observation
is that it includes the Virasoro constraint~\cite{Na1}.
Since the stochastic time evolution is generated by the noise essentially,
the emergence of Virasoro constraint
is traced to the noise correlations in eq.(5) which
are equivalent to the insertion of matrices
into the loop variable,
$
M \rightarrow M + \v\t M^{m-1}           \ ,
$
in $\phi_n$. It generates the transformation
$
[ - \v\t L_{m-2} , \phi_n ] = n \v\t \phi_{m+n-2}   \ ,
$
which corresponds to the noise correlation (5).
In fact,
for real symmetric matrix models ( non-orientable strings ),
$L_n \equiv - {N^2} X_{n+2}$
satisfies the Virasoro algebra without central extension,
\EQ
[ L_m, L_n ] = (m-n)L_{m+n}      .
\EN

It is also worthwhile to note that
the F-P equation (12) realizes the Virasoro constraint for the
probability distribution. Namely,
${\tilde L}_n \equiv {N^2} {\tilde X}_{n+2}$
also satisfies the Virasoro algebra without central extension (14).
Therefore,
the F-P equation deduces a constraint equation for the
distribution function even at the
discretized level, justifying the generation of the partition function
which satisfies the Virasoro constraint at the infinite stochastic time.
\EQ
{\tilde L}_n \lim_{\t \rightarrow \infty}P(\phi, \t)  = 0
\ , \quad {\rm for } \
n = -1, 0, 1, ...                  \ . \nn
\EN
For hermitian matrix models, the Virasoro constraint
for the partition function (15)
was found as the S-D equation~\cite{AJM}.
In the continuum limit, it deduces
the continuum version of the Virasoro
constraints~\cite{FKN}.
The expressions (8) and (11) also give a constraint on possible initial
condition dependence of the expectation value and the partition
function at the infinite stochastic time limit, such as,
$
\lim_{\t \rightarrow \infty}H_{FP}[\phi(0), {\pa\over\pa \phi(0)}]
P[\phi, \t] = 0        \ .
$
This implies that these quantities may have the initial
value dependence up to the solution of the Virasoro constraint.

\vs{8}
\centerline{\bf{Continuum Limit and Continuum Field Theory of
Non-Orientable Strings}}

Now we take the continuum limit. First we introduce a length scale
\lq\lq $a$ " and define the physical length of the loop created by
$\phi_n$ with
$
l = n a
$. Then we may redefine field variables and the stochastic time
at the continuum limit as follows.
\bea
G_{st}
&\equiv& N^{-2} a^{- D}       \  ,\nn
d\t
&\equiv& a^{- 2 + D/2} \v\t       \  ,\nn
\Phi (l)
&\equiv& a^{- D/2 } \phi_n      \ , \nn
\Pi (l)
&\equiv& a^{- 1 + D/2 } \pi_n      \ ,\nn
\ena
where we would like to keep the string coupling constant, $G_{st}$, finite at
the double scaling limit~\cite{DS}.
For the existence of the smooth
limit from the discretized stochastic time evolution to the \lq\lq
continuum" one, we require the condition, ${D\over 2} - 2 > 0$.
The scaling dimensions of all the quantities in (16)
have been
determined except the scaling dimension of the string coupling
constant, $D$,
by assuming~\cite{IK}\cite{JR},
\bea
\v\t H_{FP}
&=& d\t {\cal H}_{FP}          \  ,  \nn
\big[ \Pi(l) ,  \Phi(l)  \big]
&=& \d (l - l')                \  . \nn
\ena
Then we obtain the continuum F-P hamiltonian, ${\cal H}_{FP}$, from
$H_{FP}$ at the continuum limit,
\bea
{\cal H}^{non-or.}_{FP}
&=& - {1\over 2}\int_0^{\infty}\!dl
\big\{ 2 G_{st}\int_0^{\infty}\!dl' \Phi(l+l')l'\Pi(l')l\Pi(l) +
\int_0^{l}\!dl' \Phi(l-l')\Phi(l') l\Pi(l)                   \nn
&{}& \quad + \sqrt{G_{st}} l\Phi(l) l\Pi(l)
 + \rho(l)\Pi(l)       \big\}             \ , \nn
\ena
for non-orientable strings.
By the redefinition (16), the F-P hamiltonian (18) is uniquely fixed
at the continuum limit except the cosmological term.
To specify the explicit
form of the cosmological term $\rho(l)$ in (18),
we have to carefully evaluate the contribution which
comes from the 3-point splitting interaction term and the matrix model
potential,
$
a^{1-D}\sum_{\a=0}g_{\a}\phi_{n + \a}       \  ,
$
at the double scaling limit of the real symmetric matrix model.
Here we remember that
the S-D equation for non-orientable string at large N limit takes exactly
the same form as the
orientable one under the suitable choice of the matrix model
coupling constants.
Since the continuum limit is taken
by using the universal part of the disc amplitude~\cite{IK}, we naively expect
$\rho (l)$ takes the same form as that for orientable strings.

To show explicitly this is indeed the case,
we consider the simplest matrix potential given by,
$g_0 = -1/2,\ g_1 = g/2, g_2 = g_3 =... = 0$ in (1),
which corresponds to $c=0$.
Let us introduce the string field variable
\bea
\phi (z)
&\equiv& \sum_n z^{-1-n}\phi_n
= {1\over N}{\rm tr}{1\over z - M N^{-1/2}}           \ , \nn
\v\zeta (z)
&\equiv& \sum_n z^{-1-n} \v\zeta_n                       \ . \nn
\ena
To take the continuum limit, we redefine the field variable,
\bea
\phi (z) &\equiv& {1\over 2}(z - g z^2 )
+ c_0 z_c^{-1} a^{3/2}\Phi(u)    \ , \nn
\v\zeta (z) &\equiv& c_0 z_c^{-1} a^{3/2} d{{\tilde \zeta}} (u)
\ , \nn
\ena
where we have introduced the \lq\lq renormalized" parameters,
$
z \equiv z_c {\rm e }^{a u}       \  ,
$
and
$
g \equiv g_c {\rm e}^{ - c_1 a^2  t }    \  ,
$
and the \lq\lq continuum" stochastic time
$
d\t \equiv c_0 z_c^{-2} a^{1/2} \v\t                \ ,
$
where $z_c = 3^{1\over 4}(3^{1\over 2} + 1)$ and
$g_c = {1\over 2 \cdot 3^{3\over 4}}$ are the critical values
and $t$ denotes the cosmological
constant. The constants $c_0$ and $c_1$
are chosen for convenience.
The scaling dimension of all the quantities have been
determined so that the string coupling,
$
1/G_{st} \equiv c_0^2 N^2 a^5
$
, is fixed at the double scaling limit~\cite{DS}.

By using the Laplace transformation,
\EQ
\Phi(u) = \int_{0}^{\infty}\!dl{\rm e}^{-u l}\Phi(l)    \ ,
\EN
in the Langevin equation derived for the redefined field variable
$\Phi(u)$ in (20) with the same procedure as shown in (4),
we obtain the following Langevin
equation which is equivalent to the continuum
F-P hamiltonian (18)~\cite{Na1},
\bea
d\Phi(l)
&=& d\zeta (l)
+ {1\over 2}d\t \big\{ l \int_{0}^{l}\!dl' \Phi(l') \Phi (l-l')
+ \rho (l) + \sqrt{G_{st}} l^2 \Phi(l)
\big\}                             \ , \nn
<d\zeta (l) d\zeta (l')>
&=& 2 d\t G_{st}
l l' < \Phi(l + l') >              \ , \nn
\rho(l)
&=& 3 \d''(l) - {3 t\over 4}\d (l)         \ , \nn
\ena
for non-orientable string. It is consistent with the
naive continuum limit of its discretized version (4) except
the term $\rho (l)$.
As we have shown explicitly, the cosmological term
$\rho (l)$ takes the same form both for orientable
and non-orientable strings.
The field theory of non-orientable strings
is also consistent with ref.~\cite{Wata} in transfer matrix formalism.
The double scaling limit of the real symmetric matrix model
has been studied in a
quartic potential~\cite{RS}, while our result shows
that it happens in the cubic
potential as well. We notice that the continuum
F-P hamiltonian includes the continuum Virasoro generator ${\cal L}(l)$,
\bea
{\cal L}(l)
&=& -
\big\{ \int_0^{\infty}\!dl' \Phi(l+l')l'\Pi(l') +
{1\over 2 G_{st}}\int_0^{l}\!dl' \Phi(l-l')\Phi(l')                    \nn
&{}& \quad + {1\over 2 \sqrt{G_{st}}} l\Phi(l)
 + {1\over 2 G_{st}}{\rho(l)\over l}       \big\}             \ . \nn
\ena
These generators satisfy the continuum Virasoro algebra,
$
[{\cal L}(l), {\cal L}(l')]
= (l-l'){\cal L} (l+l')      \ .
$
In the stochastic
quantization view point, since the stochastic time scaling dimension
is given by ${D\over 2} - 2 = {1\over 2} > 0$ for $c = 0$, we expect that
the discretized version of the loop space Langevin equation
for real symmetric matrix models
may provide a possible method for numerical calculation
of non-orientable 2D random surfaces to sum up the topologies of surfaces.
In the next section, we extend the idea which
we have learned in 2D case to 4D Euclidean quantum gravity.

\vs{8}
\centerline{\bf{4D Quantum Gravity from Stochastic 3-Geometries}}
Here we point out that the time evolution
in 4D quantum gravity
is described by a Langevin equation for 3-geometries in terms of
the Ashtekar's canonical field variables by showing that the corresponding
Fokker-Planck hamiltonian operator exactly
recovers the hamiltonian of 4D Euclidean
quantum gravity in
the gauge $N=1$ and $N^i = 0$\footnote{
In 2D quantum gravity, it is pointed out that this gauge fixing
recovers the time evolution in non-critical string field
theories\cite{FIKN}.
}.
The Hartle-Hawking type boundary condition
is naturally imposed in this scheme by specifying the initial probability
distribution functional. We also present the lattice regularization of
this approach which defines a lattice regularization of Ashetekar's
canonical formalism~\cite{Na3}.

At first, We propose the basic Langevin equation for 3-geometry in terms of
the Ashtekar's canonical variables~\cite{Ash} to recover the
hamiltonian of 4D quantum gravity with the corresponding
F-P hamiltonian defined latter.
The simplest form of the Langevin equation
is defined by,
\bea
\v A^a_i(x, \t)
&=&  \v\zeta^a_i(x, \t)                          \ , \nn
<\v\zeta^a_i(x, \t) \v\zeta^b_j(y, \t)>_\zeta
&=& {\kappa\over 2} \v\t \epsilon^{abc}<F^c_{ij}(x, \t)>_\zeta\d^3(x-y)
              \ ,
\ena
where $A^a_i (x, \t)$ is a SU(2) gauge field in Euclidean
Ashtekar's canonical
formalism\footnote{
We use the notation,
$
F^a_{ij} = \pa_i A^a_j - \pa_j A^a_i + \epsilon^{abc}A^b_i A^c_j  \ ,
$
and
$
D^{ac}_i = \d^{ac}\pa_i + \epsilon^{abc}A^b_i   \ .
$
The field variable $A^a_i (x)$ is $real$ in the Euclidean
Ashtekar's formalism.
}. In this note, Latin indices \lq\lq i,j,k,...\rq\rq denote
the spatial part of the spacetime coordinate indices.
While the Latin letters
\lq\lq a,b,c,...\rq\rq denote the spatial part of the internal indices.
$x, y,...$ denote spatial spacetime coordinates.
The one step time evolution is defined by
$\v A^a_i(x, \t) \equiv A^a_i(x, \t + \v\t ) - A^a_i(x, \t)$ in (24).
The coupling constant $\kappa$ is defined by
$\kappa \equiv 16\pi G$ with $G$, the gravitational ( Newton's ) constant
in the natural unit $\hbar = c = 1$.
The noise variable in (24) is not a simple white noise.
The expectation value of the R.H.S. of
the noise correlation is understood to be taken with respect to the noises
up to the one unit time step before, $\t - \v\t$, in the sense of
Ito's stochastic calculus.
It is also equivalent to require $< \v\zeta^a_i (\t) > = 0$ in Ito's
calculus.

The invariant property of (24) is not apparent even if we introduce
algebra-valued 1-form,
$
A(x, \t) \equiv A^a_i T^a dx^i              \ ,
\v\zeta (x, \t) \equiv \v\zeta^a_i T^a dx^i  \ ,
$
and algebra-valued curvature 2-form,
$
F(x, \t) = dA + A{\wedge}A        \ .
$
Then (24) is rewritten by,
\bea
\v A (x, \t)
&=& \v\zeta (x, \t)        \ , \nn
<\v\zeta (x, \t)\wedge \v\zeta (x, \t)>
&=& \kappa\v\t \d^3 (0) F(x, \t )   \ .
\ena
The R.H.S. of the noise correlation should be regularized in a gauge
invariant way.
The basic Langevin equation is manifestly covariant under the
SU(2) local gauge transformation,
$A^a_i (x) \rightarrow A^a_i (x) + D^{ab}_i \o^b (x) \ $. While it is
not covariant under the spatial general coordinate transformation,
$A^a_i (x) \rightarrow A^a_i (x) + F^a_{ji}\xi^j (x) \ $, due to
the appearance of the divergent term,
$\kappa \v\t \d^3(0) F^a_{ji}\xi^j$, in the transformation of
the R.H.S of (24). It is formally cancelled by adding a term in the
R.H.S. of the Langevin equation (24) which
comes from the invariant path-integral measure we discuss latter.

In terms of the solution of the Langevin equation, the following
equality holds.
\bea
<\Pi_{x,i,a}\d \big( A^a_{i\ \zeta} (x, \t)
&-& A^{a(final)}_i (x)\big) >_\zeta                      \nn
&=& <A^{(final)}|{\rm e}^{-\t \tHFP [\hp, \hA ]}|A^{(initial)}>     \ , \nn
&=& <A^{(initial)}|{\rm e}^{-\t \HFP [\hA, \hp ]}|A^{(final)}>         \ .
\ena
In the L.H.S., $A^a_{i \zeta} (x, \t)$ denotes the solution of
the Langevin equation with the initial condition,
$
A^a_i (x, 0) = A^{a(initial)}_i (x)         \ .
$
In the R.H.S., $\tHFP [\hp, \hA]$ and $\HFP[\hA, \hp]$ are defined by,
\bea
\tHFP [ \hp, \hA ]
&=& - {G_0\over 4} \itd \epsilon^{abc} \hp_a^i (x) \hp_b^j (x)
 {\hat F}^c_{ij}(x)             \ ,  \nn
\HFP [\hA, \hp ]
&=& - {G_0\over 4} \itd
\epsilon^{abc} {\hat F}^c_{ij}(x)
    \hp_a^i (x) \hp_b^j (x)            \ .
\ena
To show the equality (26), the commutation relation,
\EQ
[ \hp_a^i (x) \ , \hA^b_j (y) ]
= \d_a^b \d^i_j\d^3 (x-y)              \ ,
\EN
and the vacuum, $|0>$ with
$\hp^i_a |0> = <0|\hA^b_j = 0$, have been
assumed ( see also eq.(8))~\cite{Na1}.
The Fokker-Planck hamiltonians (27) are just the hamiltonian
for 4D quantum gravity without cosmological
term in Ashtekar's variables~\cite{Ash} with different operator
orderings.
The operator $\hp_a^i$
is interpreted as the
\lq\lq triad \rq\rq in the Euclidean Ashtekar's formalism.
Namely\footnote{
$
{\tilde e}_a^i \equiv q^{1/2}e_a^i   \ ,
$
where
$
q = {\rm det}(q_{ij})   \
$
with spatial metric
$
q^{ij} = e_a^i e_a^j   \ .
$
},
$
\hp^j_a (x) = {2 i\over \kappa}{\tilde e}^j_a (x)  \ .
$
Stochastic time evolution with the Langevin equation (24)
corresponds to the Euclidean time evolution in 4D quantum gravity
with gauge fixing, $N=1$ and $N^i = 0$. In the stochastic process (24),
the \lq\lq triad \rq\rq plays the role of the noise variable and the
scale of the fluctuation is characterized by the curvature.
Though the two expressions in (26) are precisely equivalent,
the second
definition of the F-P hamiltonian in (27) has an advantage
for further discussion.
This is because the vacuum satisfies the ( local ) hamiltonian
constraint with the operator ordering in $\HFP$,
$
<0| {\cal H}(\hA (x), \hp (x)) = {\cal H}(\hA (x), \hp (x)) |0> = 0   \ ,
$
where
$
H_{FP}[\hA, \hp] \equiv \itd {\cal H}(\hA (x), \hp (x))  \ .
$
It should be noticed that the
operator
ordering of the
F-P hamiltonian operator $\HFP$ is deferent from that appeared in the F-P
equation.

Let us consider the initial distribution dependence of the probability
distribution functional by averaging the expectation value
(26) with respect to the initial probability distribution. It is defined
by integrating out the initial configuration $A^{a(initial)}_i (x)$,
on which the solution
of the Langevin equation $A^a_{i \zeta} (x, \t)$ depends,
with the distribution, $P[A^{(initial)}, 0]$, in the L.H.S. of (26).
It gives a generalized form of the distribution
functional $P[A, \t]$, which is defined by
$<O[A_\zeta] (\t)>_\zeta \equiv \igr O[A]P[A, \t]$, as follows
\bea
P[A, \t]
= \int\!{\cal D}A^{initial}
<A^{initial}|P[\hA, 0]{\rm e}^{-\t \HFP [\hA, \hp]}|A>     \ .
\ena
For an arbitrary observable $O[A]$, the average with respect to the
initial value distribution also gives,
\bea
<O[A_\zeta (\t)]>_\zeta
= \int\!{\cal D}A^{initial}
<A^{initial}|P[\hA, 0]{\rm e}^{-\t \HFP [\hA, \hp]}O[\hA ]|0>     \ .
\ena
In the definition of the expectation value in the L.H.S. of (30),
the average is
also taken with respect to the initial values with the distribution
$P[A^{initial}, 0]$. For example,
eq.(26) is given for
$O[A] = \Pi_{x, a, i} \d \Big( A^a_i(x) - A^{a(final)}_i (x) \Big)$, with
the initial distribution,
$P[A, 0] = \Pi_{x, a, i} \d \Big( A^a_i(x) - A^{a(initial)}_i (x) \Big)$.
{}From (30), the time evolution equation for the expectation values of
observables is given by,
\bea
{d \over d\t}<O[A_\zeta (x, \t)]>_\zeta
= <H_{FP}[ A^b_j (x), {\d \over \d A^a_i(x)} ]
O[A^a_i(x)] {\big|}_{A^a_i (x) = A^a_{i{}\zeta} (x, \t)}>_\zeta   \ .
\ena

The initial condition dependence in the amplitude (30) leads a restriction on
the boundary
condition with respect to the
Euclidean time.
We notice that eq.(30) also gives a constraint for initial distribution,
\bea
{d \over d\t}<O[A_\zeta (x, \t)]>_\zeta
= \int\!{\cal D}A^{in} P[A^{in}, 0]
\HFP[A^{in}, \dpi]<A^{in}|{\rm e}^{-\t \HFP [\hA, \hp]}O[\hA]|A>     \ .
\ena
The existence of the equilibrium limit requires that
the R.H.S. in (32) should be zero at the infinite stochastic time.
A trivial solution of
this
constraint is the vanishing curvature at any spatial points,
$
P[A, 0] = \Pi_{x, a, ij} \d \Big( F^a_{ij}(x) \Big)        \ .
$
This initial distribution, however, should be excluded
because the R.H.S. of (31) is identically
zero even at finite
stochastic time and there is no time development.
In general, if we choose a solution of
the hamiltonian constraint as
the initial value distribution, obviously there is no time evolution in the
corresponding F-P equation.
Especially, the constraint does not allow us to solve the Langevin
equation with the initial condition $A^{a(initial)}_i (x) = 0$.

To specify
the physical boundary condition and
define a class of
solutions for the local hamiltonian constraint of 4D quantum gravity in this
context, we may choose the following initial condition which generates a
nontrivial time evolution as an analogue in 2D case,
\bea
P_{H-H}[A, 0] = \Pi_{x \neq z_0, a, i} \d \Big( F^a_{ij}(x) )
\Big)        \  .
\ena
The spatial coordinates, $x=z_0$, is identified to the point where the
3D manifold is absorbed into nothing
in the sense of Hartle-Hawking type boundary condition. The local
hamiltonian constraint is broken at this point but it may be recovered at an
equilibrium limit.

Let us consider the gauge invariant
path-integral measure in the Ashtekar's variable, which are
implicitly assumed in the expectation value (30).
One way to specify the path-integral measure
is to introduce a regularization for the noise correlation (24).
In the following, we show that an extra term is
necessary in a lattice regularization of the Langevin equation (24)
to define an invariant and well-defined
measure by identifying the noise
correlation in (24) to a \lq\lq superspace \rq\rq
metric with Ashtekar's variable.

By using the lattice
regularization of the Langevin equation and the noise correlation (24)
, the invariant
property of the Langevin equation in the sense of Ito's calculus
naturally introduces the path-integral measure in the configuration
space with Ashtekar variables.
It is given by~\cite{Na3}
\bea
\v U(x, i)_{\a\b} (\t ) = \v W(x, i)_{\a\b} (\t ) +
\v\t |{\cal G}|^{1/2} {\pa \over \pa U(y, j)_{\c\d}}
\Big( |{\cal G}|^{-1/2} {\cal G}[x, i;y ,j]_{\a\b; \c\d} \Big)     \ .
\ena
The regularized Langevin equation describes the one step time
evolution of the link variables\footnote{
For the detailed discussion to
keep the dynamical variable within the element of SU(2) group along the time
development and for the derivation of the corresponding
lattice regularized F-P hamiltonian, see Ref.\cite{Na3}
},
$
U(x, i)(\t + \v\t ) = U(x, i)(\t ) + \v U(x, i) (\t )       \ .
$
The dynamical variable and the noise variable,
$U(x, i)$ and $\v W(x, i)$ respectively, have been assigned
on the link of 3-dimensional lattice, which is specified by the site $x$
and its nearest neighbor in the i-th direction denoted by $x + {\hat i}$.
$U(x, i)$ is an element of SU(2) group in the adjoint
representation, while noise $\v W(x, i)$ is algebra valued. The quantity
$
{\cal G}[x, i;y ,j]_{\a\b;\c\d} \ ,
$
is interpreted as the inverse of the \lq\lq superspace \rq\rq metric and
$| {\cal G}|$ denotes its determinant. The superspace is spanned by the
configuration $\{U_{\a\b}(x, i) \}$.
The inverse of the superspace metric is given in the following
regularized noise correlation.
\bea
<\v W(x,i)_{\a\b}(\t) \v W(y,j)_{\c\d} (\t)>
= \v\t <{\cal G}[x, i;y ,j]_{\a\b;\c\d}(\t)>      \ ,
\ena
where,
\bea
{\cal G}[x, i;y ,j]_{\a\b;\c\d}
&=& \d_{x+\hi, y+\hj}\d_{\b\d} \big\{
U(x,i)U(x+\hi, -j) - U(x, -j)U(x-\hj, i)   \big\}_{\a\c}      \ \nn
&+& \d_{x+\hi, y}\d_{\b\c} \big\{
U(x,i)U(x+\hi, j) - U(x, j)U(x+\hj, i)   \big\}_{\a\d}      \ \nn
&+& \d_{x, y+\hj}\d_{\a\d} \big\{
U(x+\hi, -i)U(x, -j) - U(x+\hi, -j)U(x+\hi-\hj, -i)   \big\}_{\b\c}      \ \nn
&+& \d_{x, y}\d_{\a\c} \big\{
U(x+\hi,-i)U(x,j) - U(x+\hi, j)U(x+\hi+\hj, -i)   \big\}_{\b\d}      \ .
\ena
The regularized Langevin equation (34) is invariant under the
\lq\lq general coordinate transformation \rq\rq in superspace,
$\{ U(x, i) \}$.
\bea
U(x, i)_{\a\b} \rightarrow V(x, i)_{\a\b}[U]     \ ,
\ena
where $V_{\a\b}(x, i)$ is also an element of SU(2) group in the adjoint
representation and an arbitrary functional of $U_{\c\d}(y, j)$.
The second term in
the R.H.S. in (34) is
necessary for the invariance in Ito's calculus~\cite{Na3}. The role of
the contribution from the invariant measure was first clarified in
the case of the Langevin equation for a particle moving on a Riemann
surface~\cite{Gr} ( see also Ref.~\cite{Na2} and
the Appendix in Ref.~\cite{Na} ).
The similar argument is formally possible even in the non-regularized
version (24) if we identify the R.H.S. of
noise correlation in (24) to
the continuum superspace metric.  Therefore,
the naive continuum limit of these equations, (34) (35) and (36),
coincides with those in
eq.(24) except a divergent term with $\d^3 (0)$ which comes from
the second term in the R.H.S. of the Langevin equation (34). Though the
divergent term actually
represents the contribution from the path-integral measure
and it is necessary for the invariant property of the Langevin
equation in a formal
sense, it is not well-defined without regularization. Thus the regularized
Langevin equation (34) provides a possible basis for numerical simulation
on 4D quantum gravity.

We comment on the two important consequences of the lattice regularized
Langevin equation~\cite{Na3}. One is the corresponding
equilibrium distribution,
\bea
\limt P[U, \t] = {\cal G}^{-1/2}    \ ,
\ena
which defines a measure invariant under the superspace general
coordinate transformation (37),
\bea
{\cal D}U \equiv {\cal G}^{-1/2}\Pi_{link}dU   \ ,
\ena
in the regularized version of the expectation values (29) and (30).
The other is the Schwinger-Dyson equation in this context.
It is given by,
\bea
< |{\cal G}|^{1/2} {\pa \over \pa U(y, j)_{\c\d}}
\Big( |{\cal G}|^{-1/2} {\cal G}[x, i;y ,j]_{\a\b; \c\d} \Big)
> = 0         \ ,
\ena
at an equilibrium limit.

\vs{8}
\centerline{\bf{Conclusion}}

In conclusion, we have shown that the Langevin equation for real symmetric
matrix models written by the loop variables
defines the time evolution of non-orientable
strings which defines non-orientable 2D surfaces at both
discretized and continuum levels. The partition
function in loop space satisfies the
Virasoro constraint at the equilibrium limit in both
discretized and continuum level.

The idea has been extended to 4D Euclidean quantum gravity.
In the Langevin equation (24) and the regularized one (34),
the \lq\lq triad \rq\rq in Ashtekar's variables are
realized as noise variables, which presumably represents the
stochastic 3-geometries.
The corresponding F-P hamiltonian is equivalent to the hamiltonian
of 4D Euclidean quantum gravity in the temporal gauge.
It allows us
to interpret the stochastic time as the Euclidean proper time.
The strategy we would like to adopt here is to characterize the
3-boundaries in the 4D spacetime by using the solution for both
momentum constraint and Gauss law constraint as
the initial distribution and
the observables.
As it is clear from (30), initial distributions and
observables are the key quantities to specify 3-boundaries on a
4D spacetime in this approach.
The method developed in 2D quantum gravity in loop space indicates that the
expectation values of such quantities satisfy all the constraints in
4D quantum gravity at the equilibrium limit.
There are some candidates useful to characterize 3-boundaries,
such as the extrinsic curvature term,
3D Chern-Simons term,
topological invariants~\cite{Wi} and loop variables~\cite{Smo} for
which the present formalism can be applied.
For these observables,
the gauge fixing terms for the spatial
general coordinate invariance and local Lorentz invariance,
which are introduced as a drift force in the
Langevin equation (24) following a standard
method~\cite{Zw}\cite{Na2}, do not
change the expectation value of observables at the equilibrium
limit~\cite{Na3}.

Though there is no well-defined drift force in the Langevin
equation (24) without regularization of path-integral measure, there
is another machinery to introduce a drift force.
In the Langevin equation for observables which is equivalent
to the F-P equation for the expectation value of observables (30),
a drift force appears effectively as a direct consequence of
the fact that
Ito's stochastic calculous pick up the Jacobian factor which comes from the
change of variables from the gauge fields to observables\footnote{
The simplest example is 2D Euclidean Yang-Mills theory. It can be
defined by the same Langevin equation for SU(N) as (24) with 1-dimensional
field variables $A^a (x, \t)$ and the white noise correlation,
$
<\v\zeta^a (x, \t) \v\zeta^b (y, \t)>_\zeta
= {2\over g^2} \v\t \d^{ab} \d (x-y)      \ .
$
The noise is translated as the canonical momentum variable. Then the
Langevin equation for the expectation value of Wilson loops defines
the collective field theory for 2D Yang-Mills field.
}~\cite{Na1}\cite{Na2}. In the lattice regularized Langevin equation
(34), the contribution from the invariant measure introduces a
well-defined drift force.

The problem of the present scheme is
that the noise correlation in the basic Langevin equation (24) is not
positive definite. It would force the Langevin equation (24) to be
complex, although the field variables are real in the Euclidean
Ashtekar's formalism.
The point would be main difficulty for the
numerical analysis in this scheme.
One way to deal with the problem may be to extend the Langevin equation
(24) to a class of more general gauge fixing. It is always possible by
multiplying the noise correlation in (24) with lapse function. Then one
may chose the lapse function so that the noise correlation keeps the
value to be positive definite. It is an open question if this gauge
fixing procedure, a choice of non-trivial lapse function,
make sense in numerical simulation.
Another way may be to consider the Wick rotation of the conformal
mode. Apart from these questions,
the description with the Langevin equation has a topological feature
in the sense of Nicoli-Langevin map. Such a topological feature would
relax another difficulty, renormalizability of quantum gravity.
%
%
We hope that
the approach is useful for deeper understanding of quantum gravity.


\vs{10}
\noindent
{\it Acknowledgements}

The author would like to thank J. Ambjorn, J.Greensite, A. Krasnitz,
Y. Makeenko, H. B. Nielsen, J. L. Petersen, M. Weis and Y. Watabiki
for valuable discussions and comments and all members
in high energy group at Niels Bohr Institute for hospitality.


\end{document}